\documentclass{aastex631}
\usepackage[caption=false]{subfig}
\usepackage{multirow}
\usepackage{graphicx}
\usepackage{booktabs}
\usepackage{amsmath,amssymb}
\usepackage[T1]{fontenc}
\usepackage{threeparttable}

\shorttitle{Contributions of Cascade Radiations of $e^\pm$-production to the Early Optical Afterglows of GRBs}
\shortauthors{Xiong et al.}

\begin{document}

\title{Cascade Radiations of $e^\pm$ from $\gamma\gamma$-annihilation process as an extra component of the Early Optical/X-Ray Afterglows of Gamma-Ray Bursts}

\correspondingauthor{Xiao-Li Huang}
\email{xiaoli.huang@gznu.edu.cn}

\author{Ren-Jie Xiong}
\affiliation{School of Physics and Electronic Science, Guizhou Normal University, Guiyang 550025, People's Republic of China}

\author[0000-0002-9725-7114]{Xiao-Li Huang$^{*}$}
\affil{School of Physics and Electronic Science, Guizhou Normal University, Guiyang 550025, People's Republic of China}
\affil{Guangxi Key Laboratory for Relativistic Astrophysics, School of Physics Science and Technology, Guangxi University, Nanning 530004, People's Republic of China}

\author[0000-0002-3883-6669]{Ze-Rui Wang}
\affiliation{College of Physics and Electronic Engineering, Qilu Normal University, Jinan 250200, People's Republic of China}

\begin{abstract}

Chromatic break and/or plateau observed in the early optical and X-ray afterglow lightcurves challenge the conventional external shock models of gamma-ray bursts (GRBs). Detection of TeV gamma-ray afterglows indicates strong gamma-ray production within the afterglow jets. 
We investigate the cascade radiations of the $e^\pm$ production via the $\gamma\gamma$ interaction in the jets. Our numerical calculations show that the cascade synchrotron emission can make a significant contribution to the early optical/X-ray afterglows. The combination of the primary and cascade emission fluxes can shape a chromatic break and/or plateau in the early optical/X-ray lightcurves, depending on the jet properties. Applying our model to GRBs 050801 and 080310, we found that their optical plateaus and the late X-ray/optical lightcurves can be explained with our model in reasonable parameter values. We suggest that such a chromatic optical plateau could be a signature of strong $e^\pm$ production in GRB afterglow jets. The TeV gamma-ray flux of such kind GRBs should be significantly reduced, hence tends to be detectable for those GRBs that have a single power-law decaying optical afterglow lightcurve. 
\end{abstract}


\section{Introduction}
\label{sec:intro}

Gamma-ray bursts (GRBs) are bright gamma-ray flashes from ultra-relativistic jets powered by collapses of massive stars or mergers of compact objects \citep{2004RvMP...76.1143P, 2015PhR...561....1K}. Their prompt gamma-ray emission is generally interpreted as cooling of relativistic electrons accelerated via internal shocks  \citep{1994ApJ...430L..93R, 1997ApJ...490...92K, 1998MNRAS.296..275D, 2005ApJ...628..847R, 2006ApJ...642..995P} or via internal magnetic processes within their ejecta \citep{1992Natur.357..472U, 1994MNRAS.270..480T, 2006A&A...450..887G, 2011ApJ...726...90Z}. Multiple wavelength afterglow observations are usually interpreted as cooling of electrons accelerated in the external shocks when the fireball propagates into the ambient medium \citep{1998ApJ...497L..17S}. Benefiting from the promptly-slewing capacity and precise localization as well as high sensitivity of the X-ray telescope (XRT) on board the Swift satellite, together with extensive optical follow-up observations with ground-based optical telescopes, well-sampled early X-ray and optical afterglow lightcurves were obtained for a large sample of GRBs triggered by the Burst Alert Telescope (BAT). Comprehensive analysis of the optical and X-ray afterglow data show that the early optical and X-ray afterglows are composed of different emission components \citep{2006ApJ...642..354Z,  2006ApJ...638L..67L, 2006A&A...451..821N, 2007ApJ...670..565L, 2008MNRAS.387..497P, 2011MNRAS.414.3537P, 2010ApJ...720.1513K, 2011ApJ...734...96K,2012ApJ...758...27L, 2013ApJ...774..132W, 2013ApJ...774...13L}, making the conventional external shock models cannot well explain the data \citep{2008ApJ...675..528L, 2015ApJS..219....9W}. 

In the framework of the standard external shock models, a shallow decay segment or a plateau of the early optical and X-ray afterglow lightcurves is conventionally explained as the refreshed shocks due to late energy injection from magnetic dipole radiations \citep{1998A&A...333L..87D, 2001ApJ...552L..35Z, 2018MNRAS.480.4402L, 2021ApJ...911...76X}, from the accretion of a black hole (BH) system \citep{2009ApJ...700.1047C, 2011ApJ...734...35C, 2017ApJ...849...47L}, or other origins \citep{2005A&A...443..841D, 2007ApJ...665..599T, 2014MNRAS.437.2448L, 2020ApJ...893...88O}. This scenario predicts an achromatic break in the X-ray and optical afterglow lightcurves when the energy inject ends. The observed early plateau with chromatic breaks in the optical and X-ray lightcurves challenge this scenario. In the X-ray band, the shallow decay segment is observed in almost half of the well-sampled XRT lightcurves. More interestingly, an internal plateau, which is followed by an extremely sharp drop, is occasionally observed in the X-ray afterglows of some GRBs \citep{2007ApJ...670..565L, 2007ApJ...665..599T}. Such a plateau cannot be explained with the external shock models. It is suggested that the internal X-ray plateau is contributed by the emission from the dipole radiation wind of a newborn magnetar that serves as the central engine of the GRBs \citep{2001ApJ...552L..35Z, 2007ApJ...665..599T, 2018MNRAS.480.4402L, 2021ApJ...911...76X}. On the other band, a long plateau is also observed in the early optical afterglow lightcurves of some GRBs \citep{2012ApJ...758...27L}, which is usually followed by a normal decay segment as predicted by the external shock models. In other words, the observed optical plateau maybe suggest a different physical origin. \cite{2008MNRAS.387..497P} studied the morphology of optical afterglow light curves and classified them into two prominent classes: afterglows showing early peaks and afterglows showing extended plateaus. They suggested that the two kind behaviours are resulted from the observer being located initially outside the jet aperture (for peaked afterglows) and from outflows having a non-uniform angular distribution of the ejecta kinetic energy per solid angle (for afterglows with plateaus). \cite{2011MNRAS.414.3537P} showed several afterglows of the plateaus, and suggested that the optical plateau originates from a long-lived engine. The break in the optical lightcurve marks the onset of the entire outflow deceleration. The physical origin of the plateau is indeed plausible. 

More recently, very high energy (VHE) gamma-ray afterglows are detected for several GRBs, including GRBs 180720B, 190114C, 190829A, 201015A, 201216C, 221009A, and 230307A. It is generally believed that the VHE afterglows of GRBs are attributed to the synchrotron, synchrotron self-Compton (SSC), and/or external inverse-Compton (EIC) radiations of the electrons accelerated in the jets (e.g., \citealp{2000ApJ...537..785D, 2001ApJ...559..110Z, 2001ApJ...548..787S, 2019Natur.575..459M, 2021ApJ...908...90A, 2021Sci...372.1081H, 2021ApJ...917...95Z}). The broad-band spectral energy distribution (SEDs) of these GRBs in the optical, X-ray, and sub-TeV gamma-ray bands can be roughly modeled with the synchrotron radiations and SSC process of the ultra-relativistic electrons in the jet (e.g., \citealp{2019ApJ...884..117W,2020ApJ...903L..26H, 2020MNRAS.496..974Z, 2021MNRAS.505.1718J, 2022MNRAS.512.2142Y}), although the hard VHE gamma-ray spectrum of GRB 190829A challenges the SSC scenario \citep{2021Sci...372.1081H, 2022ApJ...929...70S, 2023ApJ...947...84H}. Detection of VHE gamma-ray afterglow convincingly indicate strong VHE gamma-ray production within the jet. It was suggested that the cascade emission of electrons from the $e^{\pm}$ production process can make a considerable contribution to the early optical and X-ray afterglow \citep{2021ApJ...908..225H, 2022ApJ...939...39W}. That may be a potential explanation for the observed chromatic behaviors of the optical and X-ray afterglow lightcurves.

Motivated by the detection of TeV gamma-ray afterglows, we investigate the cascade radiations of the $e^\pm$ production via the $\gamma\gamma$ interaction in the Jets. We explore whether the chromatic break and/or plateau observed in the early optical and X-ray afterglow lightcurves are attributed to the cascade radiation. This Letter is organized as follows. Section \ref{sec:model} describes our model, then Section \ref{sec:numerical} presents the numerical results and the cases study for GRBs 050801 and 080310. conclusions and discussion are presented in Section \ref{sec:discussion}.

\section{Models}
\label{sec:model}

Both the synchrotron radiations and SSC process of the primary electron population accelerated within the jet and the cascade electron population via the $e^{\pm}$ production are considered in our model. The energy distribution of the primary electrons is taken as a single power-law function $dN/d\gamma_{e}\propto\gamma_{e}^{-\rm p}$, where $\gamma_{e}$ is the electron Lorentz factor. The $e^{\pm}$ production happens via interaction between 
a seed photon with energy $\varepsilon_{\gamma}$ and a target photons with an energy satisfying the condition of $\epsilon_{t}\ga \Gamma^{2}(m_{e}c^{2})^{2}/\rm \epsilon_{\gamma}$ in the jet comoving frame. The secondary $e^{\pm}$ pairs then produce cascade emission through synchrotron radiation and the inverse-Compton (IC) process. If the opacity of these secondary photons $\tau_{\gamma\gamma} >> 1$, they can induce to the next annihilation phase. As a result, the energy of high-energy photons is redistributed into lower-energy photons, until the opacity of the secondary photons becomes $\tau_{\gamma\gamma} < 1 $. That is the so-called electromagnetic cascade processes. The distribution of the cascaded electron population in a quasi-steady state within an time interval $\delta t$ can be given by (\citealp{2021ApJ...908..225H}; Please see original papers \citealp{2013ApJ...768...54B,2017ApJ...847...39V}) 
\begin{equation}
\label{eq:Ne1}
N_e^{\rm sec}(\gamma_e,t + \delta t)=N_e^{\rm sec}(\gamma_e^*,t)\frac{d\gamma_e^*}{d\gamma_e} + \left\{
\begin{array}{lll}
\frac{1}{\dot{\gamma}_e}\int_{\gamma_e}^{\propto} d\widetilde{\gamma}_e\dot{N}_e^{\gamma\gamma}(\widetilde{\gamma}_e,t + \delta t), && t_{e}^{\rm cool}(\gamma_{e}) < \delta t,\\
\dot{N}_e^{\gamma\gamma}(\gamma_e,t + \delta t)\delta t, && t_{e}^{\rm cool}(\gamma_{e}) > \delta t,
\end{array}\right.
\end{equation}
where $\dot{N}_{\rm e}^{\gamma\gamma}$ is the pair production rate, $\gamma_e^*$ is the electron Lorentz factor at $t$, and $t_{e}^{\rm cool}$ is the electron cooling timescale. Due to the cooling effect, the Lorentz factor will decrease from $\gamma_e^*$ to $\gamma_e$ during the time interval $\delta t$. This effect is important and redistributed the energy from high-energy SSC photons to the lower one \citep{2021ApJ...908..225H}. Note that the pair productions for the high energy photons from both the synchrotron radiations and the SSC process are taken into account in our model. The Klein-Nishina (KN) effect is also considered \citep{2005MNRAS.363..954M, 2009ApJ...703..675N, 2010ApJ...712.1232W}. The dynamic evolution of the jet is taken as the standard forward shock model \citep{1998ApJ...497L..17S,1999MNRAS.309..513H, 2000ApJ...543...66P, 2001ApJ...548..787S, 2013NewAR..57..141G, 2015PhR...561....1K}. 

\section{Numerical Results and Case Study}
\label{sec:numerical}

\subsection{Numerical calculation}

Following \cite{2021ApJ...908..225H}, we adopt a semi-analytical method to calculate the cascade emission of the $e^{\pm}$ population. To balance accuracy and efficiency of the numerical calculation, we adopt $f_{\gamma_{e}}=0.9$ \citep{2013ApJ...768...54B}, which means that the two electrons produced during the $e^{\pm}$ production process carry $90\,\%$ and $10\,\%$ of the energy of high-energy photons, respectively. In our numerical calculations for Eq. \ref{eq:Ne1}, we set the time interval as $\delta t=(10^{0.01}-1)t$ 
by dividing the time period of $0.1\,$s$-10^{8}\,$s equally into 901 bins in logarithmic space.

We numerically investigate the contributions of the $e^\pm$ cascade emission in three model parameter sets that may represent a normal GRBs in a typical density medium, a normal GRBs in a dense medium, and an extremely energetic GRBs in a typical density medium. The extragalactic background light (EBL) absorption effect is taken into account for calculating the observed high energy photons \citep{2012MNRAS.422.3189G}.  

\begin{figure}[htbp]
\centering
\includegraphics[width=0.3\textwidth, angle=0]{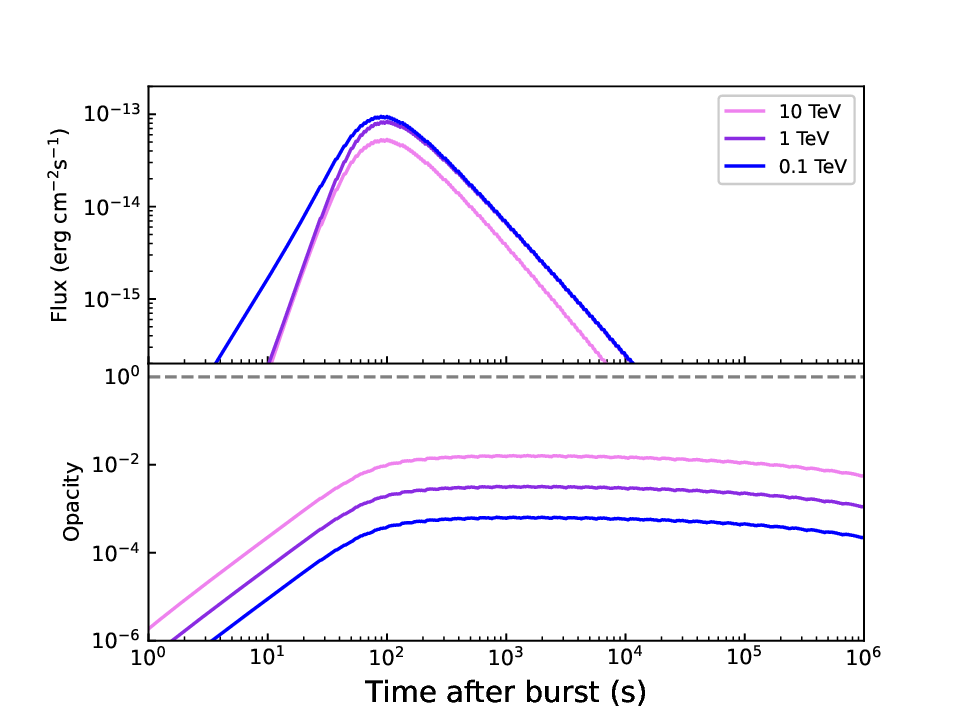}
\includegraphics[width=0.3\textwidth, angle=0]{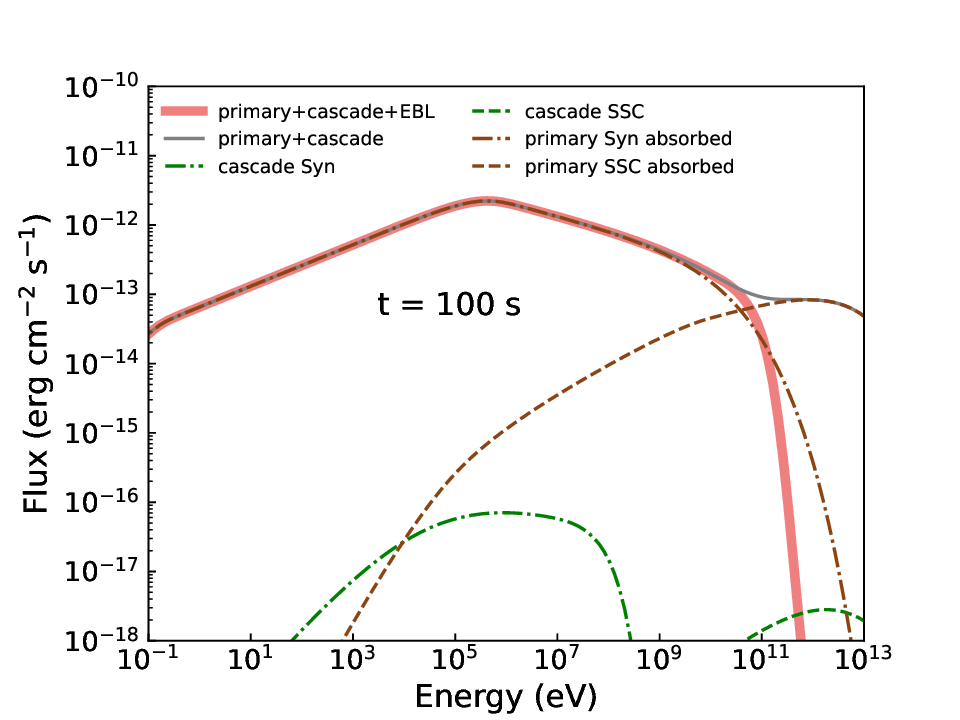}
\includegraphics[width=0.3\textwidth, angle=0]{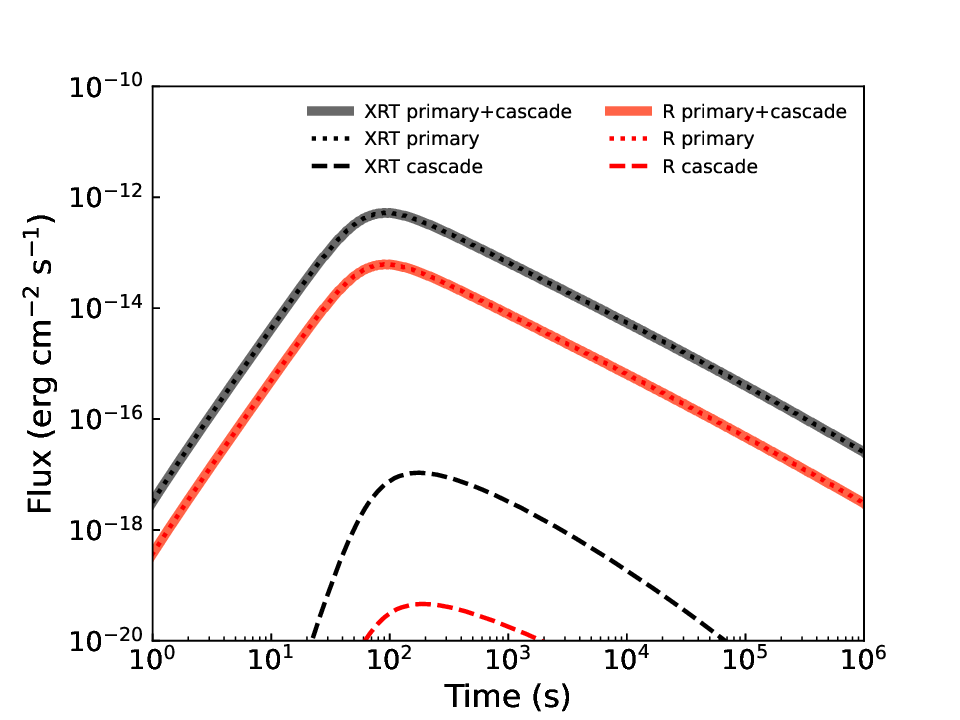}
\caption{Temporal evolution of the opacity of 0.1 TeV, 1 TeV , and 10 TeV gamma-ray photons within the afterglow jet (left panel), the broadband SEDs of afterglows at 100 seconds after the GRB trigger (middle panel), and the theoretical X-ray and optical lightcurves of the emission primary electron population and cascade emission of the $e^{\pm}$ from the $\gamma\gamma$ annihilation process (right panel). The model parameter values used are $E_{\rm k,iso} = 1 \times 10^{52} \rm \, erg$, $n_0 = 1.0 \rm \, cm^{-3}$, $\epsilon_e = 0.01$, $\epsilon_B = 1.0 \times 10^{-4}$, $p = 2.4$, $\Gamma_0 = 100$, $\theta_j = 5^{\circ}$, and $z = 1$. The line styles for different components are marked in each panels.}
\label{fig:52-1-100-0.01-opacity}
\end{figure}

\begin{figure}[htbp]
\centering
\includegraphics[width=0.3\textwidth, angle=0]{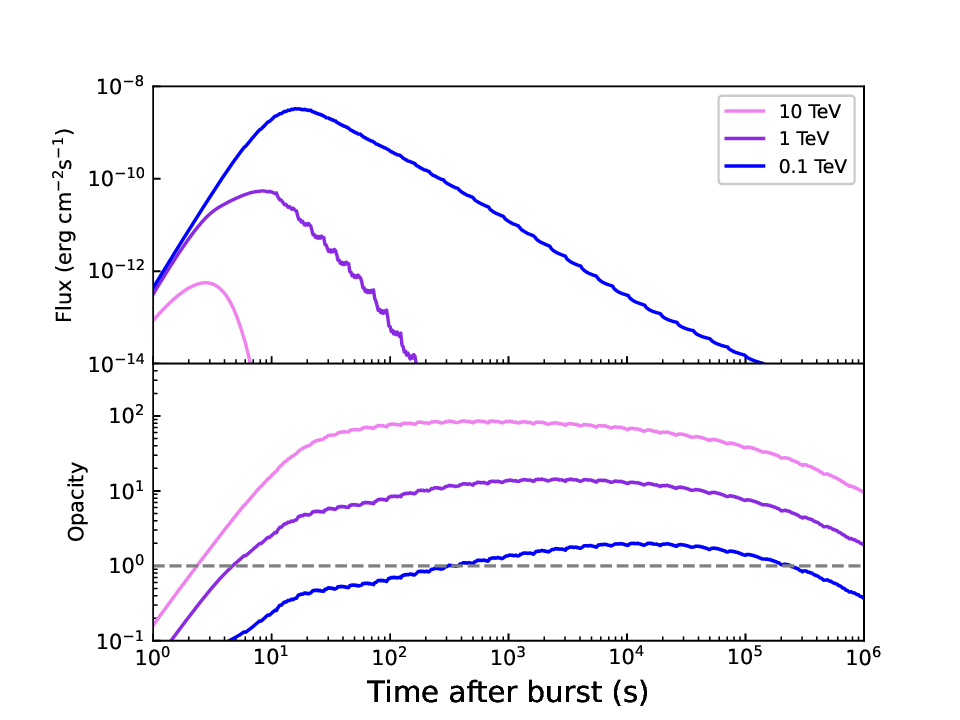}
\includegraphics[width=0.3\textwidth, angle=0]{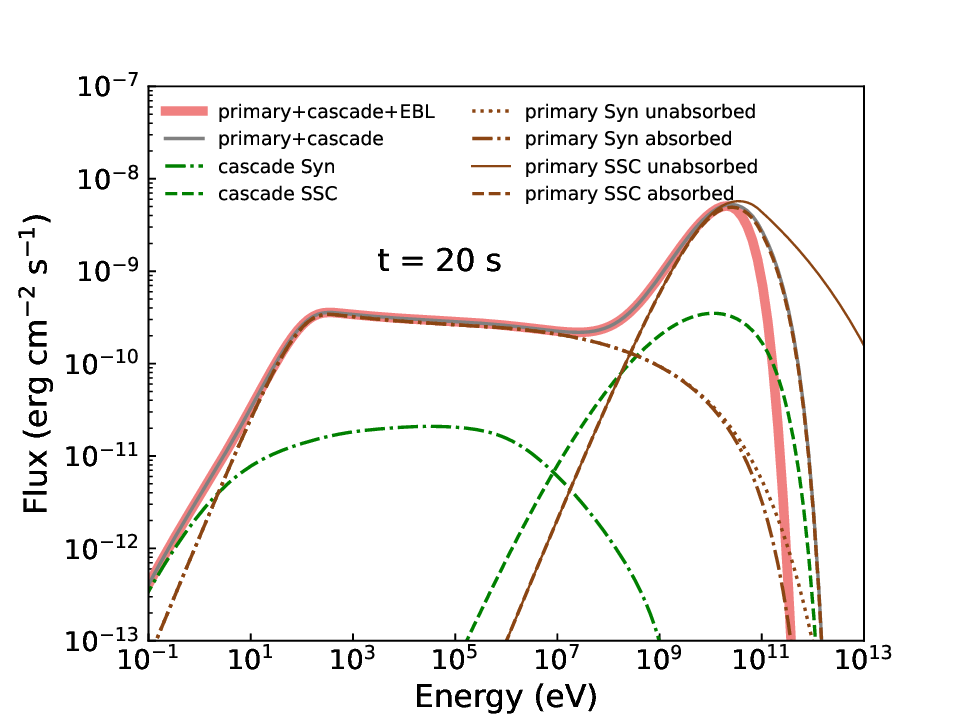}
\includegraphics[width=0.3\textwidth, angle=0]{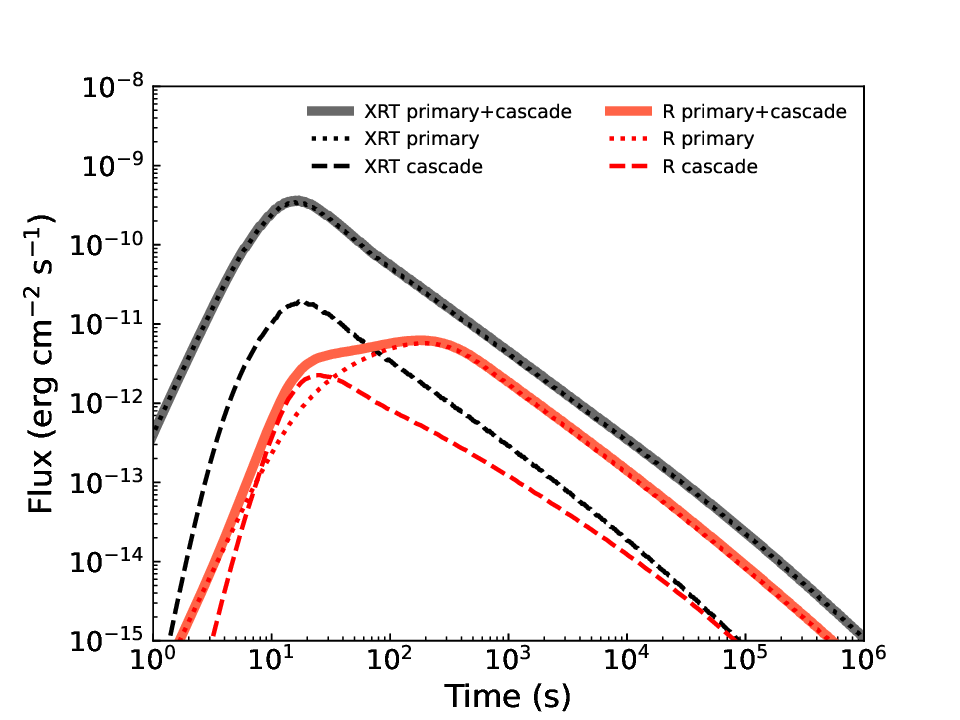}
\caption{The same as Figure \ref{fig:52-1-100-0.01-opacity}, but for $n_0 = 100.0 \rm \,cm^{-3}$ and $\epsilon_e = 0.1$.} 
\label{fig:52-100-100-0.1-opacity}
\end{figure}

\begin{figure}[htbp]
\centering
\includegraphics[width=0.3\textwidth, angle=0]{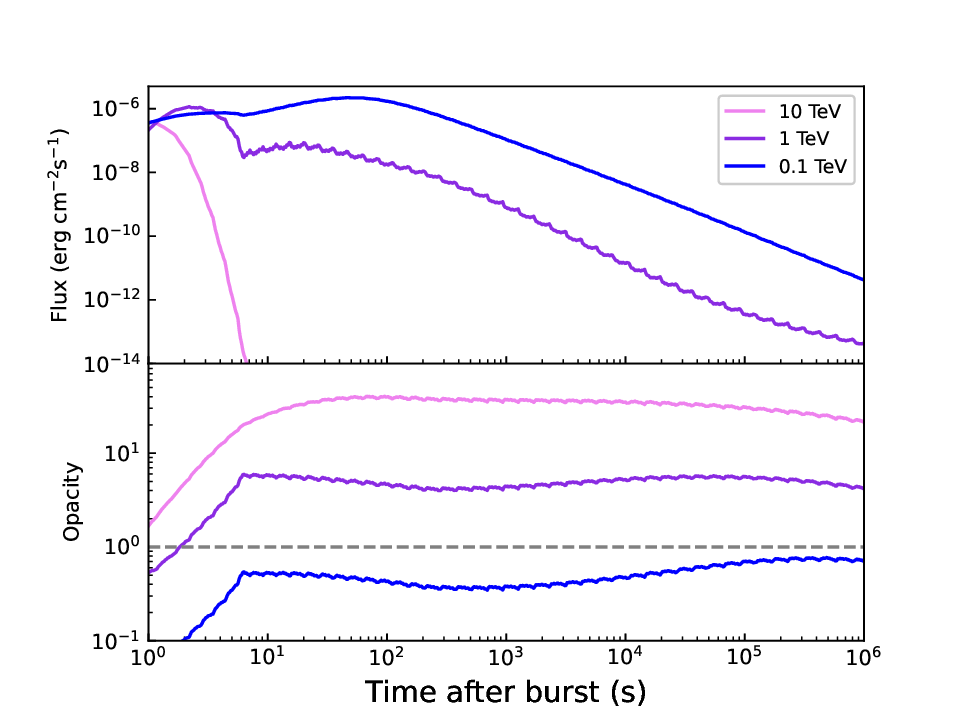}
\includegraphics[width=0.3\textwidth, angle=0]{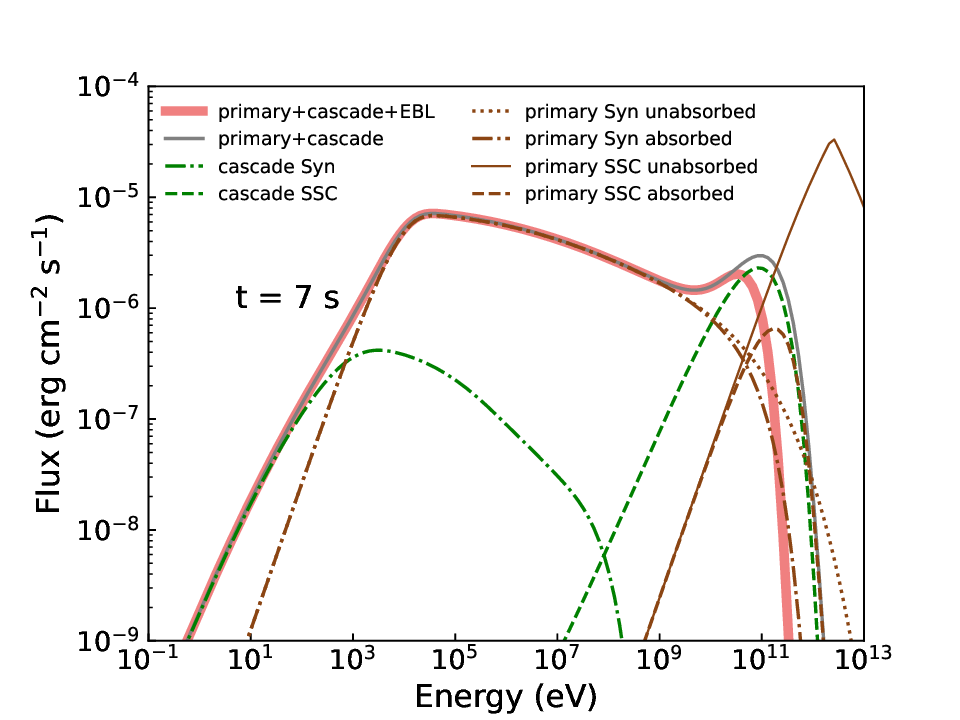}
\includegraphics[width=0.3\textwidth, angle=0]{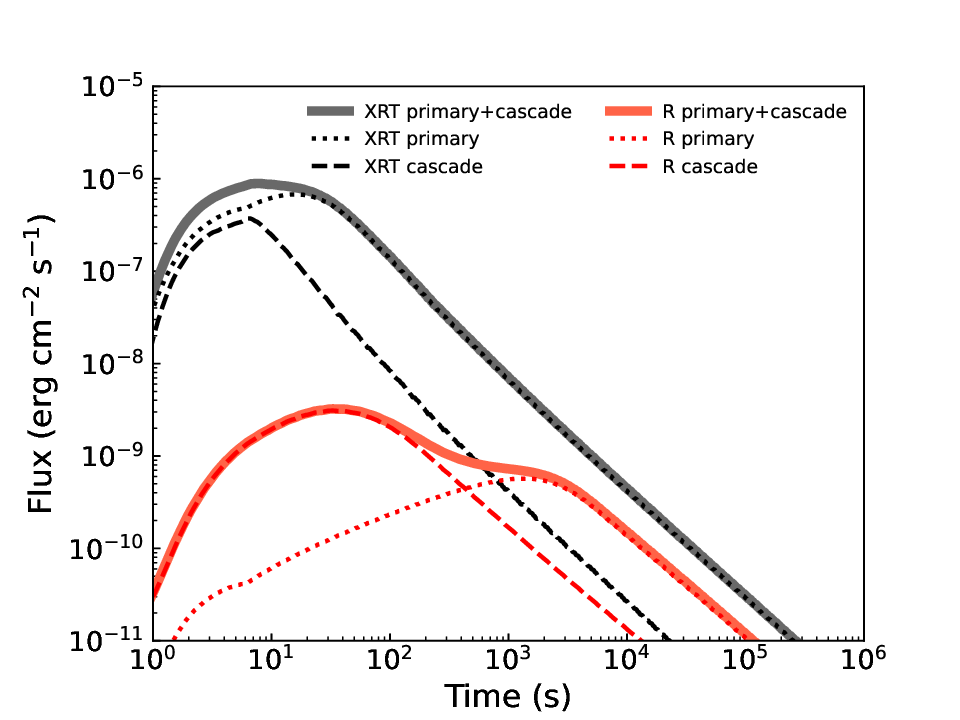}
\caption{The same as Figure \ref{fig:52-1-100-0.01-opacity}, but for 
adopting the parameter values as $E_{\rm k,iso} = 1 \times 10^{55} \rm\, erg$, $n_0 = 1.0 \rm \,cm^{-3}$, $\epsilon_e = 0.1$, $\epsilon_B = 1.0 \times 10^{-4}$, $p = 2.4$, $\Gamma_0 = 1000$, $\theta_j = 5^{\circ}$, and $z = 1$. }
\label{fig:55-1-1000-0.1-opacity}
\end{figure}

\begin{itemize}
    \item Normal GRB in a typical density medium: The parameter values are adopted as $E_{\rm k,iso} = 1 \times 10^{52} \rm \, erg$, $n_0 = 1.0 \rm \, cm^{-3}$, $\epsilon_e = 0.01$, $\epsilon_B = 1.0 \times 10^{-4}$, $p = 2.4$, $\Gamma_0 = 100$, $\theta_j = 5^{\circ}$, and $z = 1$. The 0.1, 1, and 10 TeV gamma-ray lightcurves of the primary electrons and the temporal evolution of the gamma-ray photon opacity as a function are shown in the left panel of Figure \ref{fig:52-1-100-0.01-opacity}. One can find that the peak gamma-ray fluxes at $t\sim 100$ seconds are lower than $10^{-13}$ erg cm$^{-2}$ s$^{-1}$. Their opacity is less then 0.1, indicating that they can freely escape out of the jet. The pair production rate is extremely low and cascade emission is illegible, as shown in the middle and right panels of Figure \ref{fig:52-1-100-0.01-opacity} for the spectral energy distribution (SED) at $t=100$ seconds and the X-ray/optical afterglow lightcurves of this scenario.   
    
    \item Normal GRB in a dense medium: This scenario is the same as the above one, but only changes the medium density to $n_0 =100.0$ cm$^{-3}$. As shown in Figure \ref{fig:52-100-100-0.1-opacity}, the peak flux of $0.1$ TeV gamma-ray photons is $5\times 10^{-9}$ erg cm$^{-2}$ s$^{-1}$ at $t=15$ seconds, and their opacity is $\leq 1$ at $t<500$ seconds. This indicates that bright sub-TeV afterglow emission is detectable with current telescopes, such as LHAASO (Large High Air Altitude Shower Observatory),  MAGIC (The Major Atmospheric Gamma-Ray Imaging Cherenkov), H.E.S.S., and CTA (Cerenkov Telescope Arrays). The opacity of the gamma-ray photons above 1 TeV is larger than 1. This results in strong $e^{\pm}$ production and suppression of the escape of these photons. The cascade emission makes a significant contribution to the early optical and inferred afterglow flux, as shown in the SED at $t=20$ seconds. The combination of the primary emission and cascade emission shape a plateau in the early optical afterglow lightcurve. Since the cascade emission in the X-ray and is much lower than the primary emission in this scenario, the X-ray afterglow lightcurve does not show such a plateau. 

    \item Extremely energetic GRB in a typical density medium: This is similar to GRB 221009A like extremely bright GRB. We adopt the parameter set as $E_{\rm k,iso} = 1 \times 10^{55} \rm\, erg$, $n_0 = 1.0 \rm \,cm^{-3}$, $\epsilon_e = 0.1$, $\epsilon_B = 1.0 \times 10^{-4}$, $p = 2.4$, $\Gamma_0 = 1000$, $\theta_j = 5^{\circ}$, and $z = 1$. The early gamma-ray photons emitted by the primary electron populations are firstly dominated by the synchrotron radiations then transit to the SSC process dominated phase, as shown in the middle panel of Figure \ref{fig:55-1-1000-0.1-opacity}. This makes some complex features at the early stage. In this extreme parameter, the 0.1 TeV lightcurve at the early stage is contributed by the synchrotron emission, but it is contributed by the SSC emission component after $\sim 10 \,\rm s$.  The 1 TeV lightcurve at in the early stage is dominated by the synchrotron radiations, and transits to be dominated by the SSC component at $\sim 7 \,\rm s$. The lightcurve of the 10 TeV band is from the SSC emission component. As shown in Figure \ref{fig:55-1-1000-0.1-opacity}, the 0.1 TeV gamma-rays are transparent and can freely escape out of the jet. This would be due to the high $\Gamma_0$ of the jet. Thus, bright sub-TeV gamma-ray afterglow should be detectable for such kind of GRBs. The opacity of the gamma-ray photons above 1 TeV is larger than 1. Strong $e^\pm$ process happens, hence the cascade emission makes great contributions to the early optical/X-ray afterglows. The SED at $t=7$ seconds shows that observable fluxes below $10^{3}$ eV and above $10^{10}$ eV are dominated by the the cascade emission. This dramatically modifies the shapes of the early optical, X-ray, and TeV gamma-ray afterglow lightcurves, as shown in the right panel of Figure \ref{fig:55-1-1000-0.1-opacity}. The cascade X-ray emission broadens the peak of the early X-ray afterglow lightcurve. The cascade optical afterglows at the early stage (several thousands post the GRB trigger) is much brighter than that of the primary emission component, shaping the lightcurves as a bi-peak feature. Therefore, the early optical and X-ray afterglow lightcurves are dramatically different. 

\end{itemize}

\subsection{Case Study}
\begin{figure}[htbp]
\centering
\includegraphics[width=0.45\textwidth, angle=0]{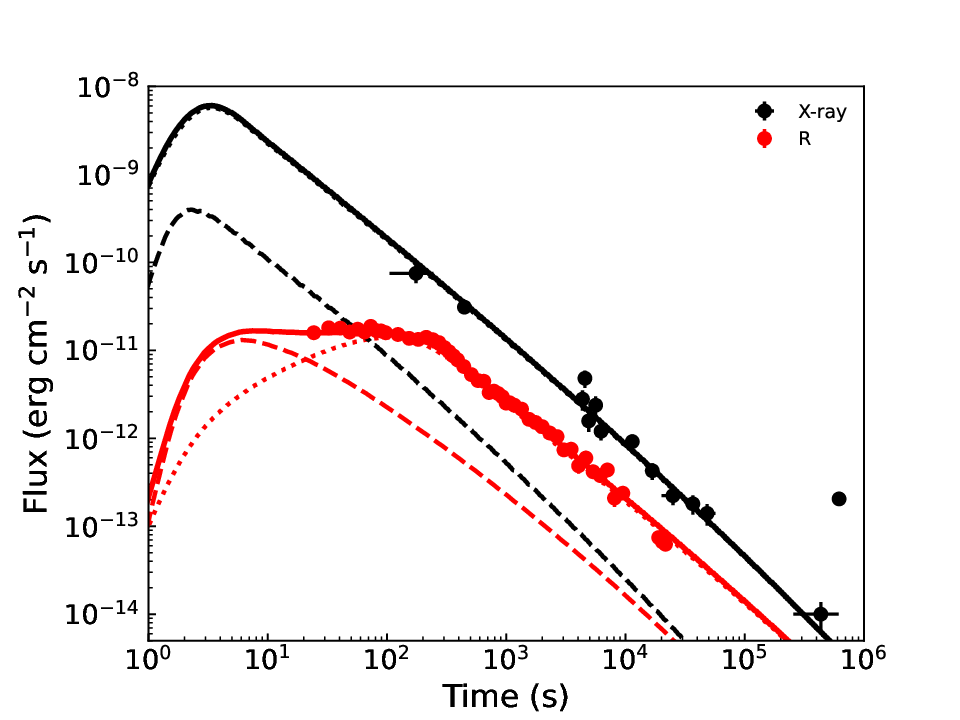}
\includegraphics[width=0.45\textwidth, angle=0]{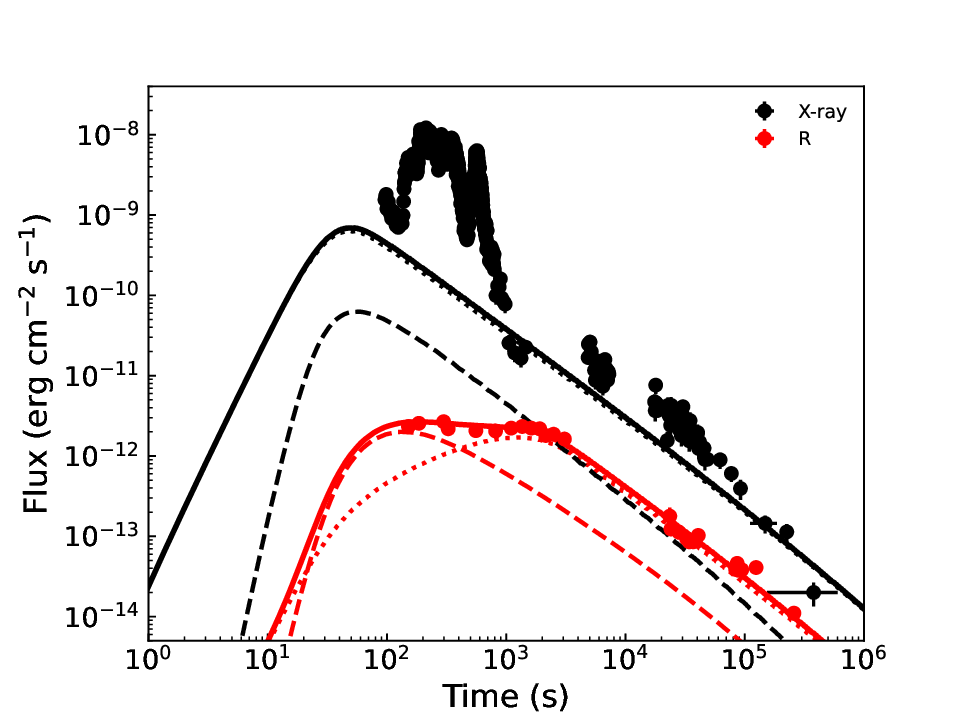}
\caption{The observed optical/X-ray lightcurves (dots) of GRB 050801 (left panel) and GRB 080310 (right panel) and the theoretical fits (solid lines) with our model. The primary and cascade emission components are marked with dotted and dashed lines, respectively.}
\label{fig:case study}
\end{figure}

The above analysis illustrates that the cascade synchrotron emission can make a significant contribution to the fluxes of the early optical/X-ray afterglows if the jet is in a dense medium or the jet is extremely energetic. They shape a chromatic break and/or plateau in the early optical/X-ray lightcurves, depending on the jet properties. As case study, we apply our model to explain the long plateau in the early optical afterglow lightcurves of GRBs 050801 and 080310. Their X-ray afterglow data are observed with the \textit{Swift}/XRT. They are taken from the Swift Burst Analyzer \citep{2010A&A...519A.102E}. Their optical data are taken from \cite{2015ApJ...805...13L}. The optical data are corrected by the extinction of our Galaxy. The optical/X-ray afterglow lightcurves are shown in Figure \ref{fig:case study}.  

\begin{itemize}
   \item GRB 050801: Its redshift is $z = 1.56$. A clear plateau is shown up in its early optical lightcurve. The X-ray afterglow lightcurve is a single power-law function without detection of a break feature. The slope of the optical lightcurve post the break is consistent with the X-ray one. We fit the optical and X-ray afterglow lightcurves with our model. Our fitting results are also shown in Figure \ref{fig:case study}. The derived model parameters are $E_{\rm k,iso} = 8.2 \times 10^{52} \rm\, erg$, $\Gamma_0 = 350$, $p = 2.4$, $\theta_j = 5^{\circ}$, $n_0 = 20.0 \rm \,cm^{-3}$, $\epsilon_B = 1.98 \times 10^{-4}$, and $\epsilon_e = 0.04$. One can find that the early optical plateau can be interpreted with the combination of the primary and cascade emission components.       

   \item GRB 080310: Its a bright GRB at $z=2.4266$. It is a typical example that have a long plateau with break at $\sim 3\times 10^4$ seconds in the optical band and at $\sim 1\times 10^3$ seconds in the X-ray band afterglow lightcurves. The decay slopes of its late optical and X-ray lightcurves post the breaks are similar. Since the X-ray plateau could attribute to the dipole radiations of the new-born magnetar, which serves as the central engine of the GRB (e.g., \citealp{2018MNRAS.480.4402L}), we fit the global optical afterglow lightcurve and late X-ray afterglow post the X-ray break with our model. Our fitting results are also shown in Figure \ref{fig:case study}. The derived model parameters are $E_{\rm k,iso} = 1.4 \times 10^{53} \rm erg$, $\Gamma_0 = 190$, $p = 2.4$, $\theta_j = 5^{\circ}$, $n_0 = 1.0 \rm\, cm^{-3}$, $\epsilon_B = 1.25 \times 10^{-4}$, and $\epsilon_e = 0.3$. One can find that the early optical plateau and late optical/X-ray afterglow lightcurves can be represented with our model. 
\end{itemize}

\section{Conclusions and Discussion}
\label{sec:discussion}

Motivating by the detection of VHE gamma-ray afterglows of GRBs, we have present an analysis of the cascade emission within the GRB jet. Our numerical calculations show that the cascade synchrotron emission can make a significant contribution to the early optical/X-ray afterglows. The combination of the primary and cascade emission fluxes can shape a chromatic break and/or plateau in the early optical/X-ray lightcurves, depending on the jet properties. Applying our model to GRBs 050801 and 080310, we found that their optical plateaus and the late X-ray/optical lightcurves can be explained with our model in reasonable parameter values. 

In the conventional external shock models, a shallow decay segment or a plateau in early afterglow lightcurves is regarded as a clear signature of the late energy injection to the afterglow fireball (e.g. \citealp{1998A&A...333L..87D, 2006ApJ...642..354Z}). As a dynamic effect, an achromatic break of this segment/plateau is expected to be observed in the optical/X-ray bands. The detection of chromatic breaks confidently rule out this scenario. \cite{2020MNRAS.492.2847B} suggested that the plateau can result from off-axis observations to structured jets. As a viewing angle effect, this scenario also predicts an achromatic onset feature of the optical and X-ray afterglows as the jet expansion before $\theta_v<1/\Gamma$, where $\theta_v$ is the light of sight to the jet axis. Our model reasonably explains the chromatic optical plateau with the cascade emission within the jet. Since the cascade emission highly depends on the model parameters, it can make the observed deviation of the optical and X-ray afterglow lightcurves in confronting with the standard external shock models. The combination of the primary and cascade emission usually flattens the early optical afterglow lightcurve. This could be a signature of the strong $e^\pm$ production within the jet. Inspecting the optical/X-ray lightcurves of GRBs with VHE gamma-ray afterglow detection \citep{2019Natur.575..455M, 2019Natur.575..459M, 2019Natur.575..464A, 2021Sci...372.1081H}, one can find that their optical afterglow lightcurves are usually a single power-law function without having a clear shallow decay segment or a plateau. This likely implies that no strong  $e^\pm$ production within the jet.  

\begin{figure}[htbp]
\centering
\includegraphics[width=0.45\textwidth, angle=0]{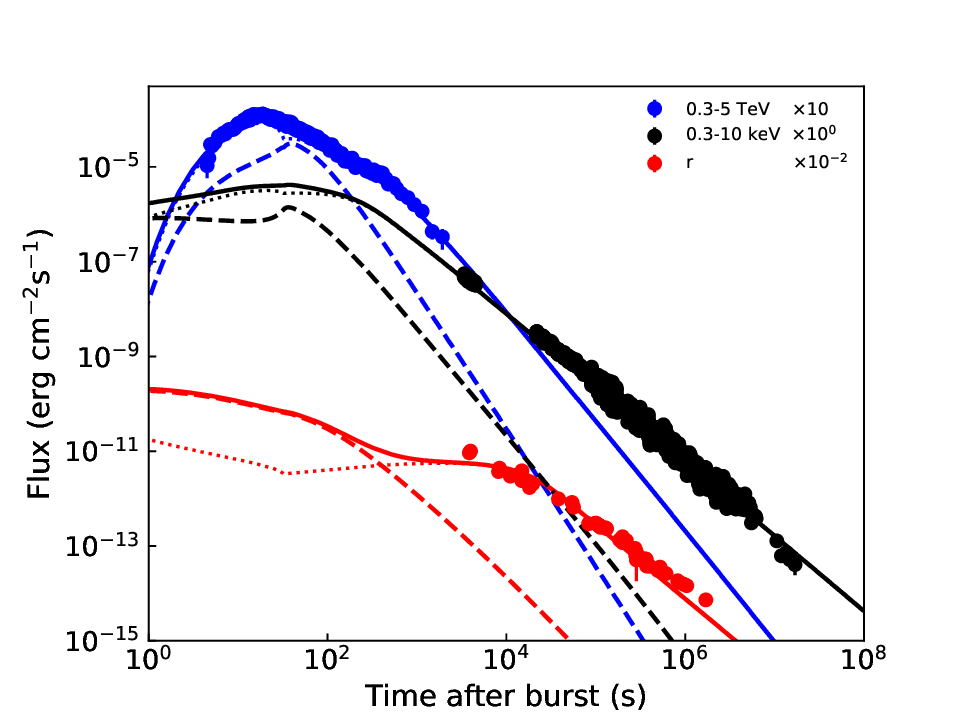}
\caption{The observed optical/X-ray/TeV lightcurves (dots) of GRB 221009A and the theoretical fits (solid lines) with our model. The primary and cascade emission components are marked with dotted and dashed lines, respectively. The model parameter values used are $E_{\rm k,iso} = 2.82 \times 10^{55} \rm \, erg$, $A_* = 0.022$, $\epsilon_e = 0.2$, $\epsilon_B = 1.0 \times 10^{-4}$, $p = 2.4$, $\Gamma_0 = 320$, $\theta_j = 0.8^{\circ}$, and $z = 0.1505$.}
\label{fig: GRB 221009A}
\end{figure}

GRB 221009 is the most energetic GRB observed so far \citep{2023Sci...380.1390L}. Its TeV afterglow emission was observed  and  photons with energy up to $\sim 10$ TeV was detected with the LHAASO telescope \citep{2023SciA....9J2778C}.  
We apply our model to fit its lightcurves in the TeV, X-ray, and optical band. The optical data are taken from \cite{2023ApJ...948L..12K} and corrected by the extinction of our Galaxy. The simultaneous X-ray data observed with \textit{Swift}/XRT are also presented. As shown in Figure \ref{fig: GRB 221009A}, the observed lightcurves in the $0.3-5 \rm\, TeV$, $0.3-10\, \rm keV$, and $r$ band are well fit with our model in the wind medium scenario. The derived model parameters are $E_{\rm k,iso} = 2.82 \times 10^{55} \rm\, erg$, $\Gamma_0 = 320$, $p = 2.4$, $\theta_j = 0.8^{\circ}$, $A_{*} = 0.022$, $\epsilon_B = 1.0 \times 10^{-4}$, and $\epsilon_e = 0.2$. One can find that our model perfectly represents the lightcurves. The cascade optical emission dominates the early ($t<100$ seconds) optical afterglows, but the X-ray and TeV afterglows are dominated by the primary emission even at the very stage. Comparing the lightcurves of the cascade emission between the scenarios of the uniform and wind mediums shown in Figure 2 and Figure 5, one can find they are dramatically different. In the scenario of the uniform medium, the cascade emission at early of the optical afterglow smoothly decays. Therefore, the early optical observations should be critical for revealing the cascade emission component and the medium feature. We should note that the feature of the cascade emission lightcurves is sensitive to the medium properties and the jet parameters \citep{2021ApJ...908..225H}.  

\begin{acknowledgements}
We thanks profs. Xiang-Yu Wang and Ruo-Yu Liu for helpful discussion. 
This work is supported by the National Natural Science Foundation of China (Grant Nos. 12203015, 12203024) and by the Department of Science $\&$ Technology of Shandong Province under Grant No. ZR2022QA071.
\end{acknowledgements}


\end{document}